\documentclass{pasj01}

\newcommand{\beq}{\begin{equation}}
\newcommand{\beqa}{\begin{eqnarray}}
\newcommand{\eeq}{\end{equation}}
\newcommand{\eeqa}{\end{eqnarray}}

\newcommand{\simgt}{\lower.5ex\hbox{$\; \buildrel > \over \sim \;$}}
\newcommand{\simlt}{\lower.5ex\hbox{$\; \buildrel < \over \sim \;$}}

\newcommand{\bd}[1]{\mbox{\boldmath $#1$}}

\begin{document}
\SetRunningHead{M. Shirasaki et al.}
{Probing cosmology with weak lensing selected clusters II}

\title{
Probing cosmology with weak lensing selected clusters II: \\
Dark energy and $f(R)$ gravity models
}


 \author{%
   Masato \textsc{shirasaki}\altaffilmark{1},
   Takashi \textsc{hamana}\altaffilmark{1}
   and
   Naoki \textsc{Yoshida}\altaffilmark{2,3,4}}
 \altaffiltext{1}{National Astronomical Observatory of Japan, 
Mitaka, Tokyo 181-8588, Japan}
 \email{masato.shirasaki@nao.ac.jp}
 \altaffiltext{2}{Department of Physics, School of Science, The University of Tokyo, 
7-3-1 Hongo, Bunkyo, Tokyo 113-0033, Japan}
 \altaffiltext{3}{Kavli Institute for the Physics and Mathematics of the Universe (WPI), 
Todai Institutes for Advanced Study, The University of Tokyo, Kashiwa, Chiba 277-8583, Japan}
 \altaffiltext{4}{CREST, Japan Science and Technology Agency, Kawaguchi, Saitama, Japan}


\maketitle

\begin{abstract}
Ongoing and future wide-field galaxy surveys can be used to
locate a number of clusters of galaxies with cosmic shear measurement alone.
We study constraints on cosmological models using statistics of
weak lensing selected galaxy clusters.
We extend our previous theoretical framework to model the statistical
properties of clusters in variants of cosmological 
models as well as in the standard $\Lambda$CDM model.
Weak lensing selection of clusters does not rely on
the conventional assumption such as the relation between luminosity and mass 
and/or hydrostatic equilibrium, but a number of observational effects
compromise robust identification.
We use a large set of realistic mock weak-lensing catalogs
as well as analytic models to perform a Fisher analysis
and make forecast for constraining two competing cosmological models,
$w$CDM model and $f(R)$ model proposed by Hu \& Sawicki,
with our lensing statistics.
We show that weak lensing selected clusters are excellent probe
of cosmology when combined with cosmic shear power spectrum
even in presence of galaxy shape noise and masked regions.
With the information of weak lensing selected clusters, 
the precision of cosmological parameter estimate
can be improved by a factor of $\sim1.6$ and $\sim8$ 
for $w$CDM model and $f(R)$ model, respectively.
Hyper Suprime-Cam survey with sky coverage of $1250$ squared degrees
can constrain the equation of state of dark energy $w_{0}$ with a level of 
$\Delta w_0 \sim0.1$. It can also constrain the additional scalar
degree of freedom in $f(R)$ model
with a level of $|f_{R0}| \sim5\times10^{-6}$, when constraints from
cosmic microwave background measurements are incorporated.
Future weak lensing surveys with sky coverage of $20,000$ squared degrees
will place tighter constraints on $w_{0}$ and $|f_{R0}|$ 
even without cosmic microwave background measurements.
\end{abstract}

\section{INTRODUCTION}
Statistical analyses of large astronomical data set have established
the standard cosmological model called $\Lambda$CDM model.
Various statistical methods have been proposed, and have been actually
applied to real observational data.
Well-known examples include 
luminosity distance to Type Ia supernovae 
\citep{2014AA...568A..22B}, 
baryon acoustic oscillation measurement with galaxy clustering 
\citep{2011MNRAS.416.3017B, 2011MNRAS.418.1707B, 2014MNRAS.441...24A},
the anisotropies of cosmic microwave background (CMB)
\citep{Hinshaw2013, 2014AA...571A..16P},
measure of gravitational growth by large-scale structure 
\citep{2012MNRAS.425..415S, 2014MNRAS.443.1065B},
and weak gravitational lensing \citep{Kilbinger2013}. 
Although the standard $\Lambda$CDM model is consistent 
with the results from a broad range of cosmological analyses,
several alternative models are still allowed at present.
In particular, the apparent 
cosmic acceleration can be explained not only by $\Lambda$CDM model
but also by a different class of cosmological models
that invoke evolving dark energy or modified gravity.
Such alternative models are basically constructed by 
the modification of Einstein equation:
\beqa
G_{\mu\nu} = 8\pi G T_{\mu \nu}. \label{eq:Einstein}
\eeqa
Dark energy models suppose an exotic form of energy 
in the right hand side of Eq.~(\ref{eq:Einstein}), 
while modified gravity models
changes the left hand side of Eq.~(\ref{eq:Einstein})
without assuming an unknown energy.
In order to test the two scenarios, 
it is essential to measure the gravitational growth of 
matter density fluctuations,
because the modification of gravity could induce characteristic 
clustering patterns to the matter density distribution in the universe.

Gravitational lensing is one of the most promising tools 
to probe matter density distribution in the universe.
The foreground gravitational field causes small image distortion
of distant galaxies.
The small distortion contains, collectively, rich cosmological 
information about the matter distribution.
Since image distortion induced by gravitational lensing is very small in general,
we require the statistical analysis in 
order to extract cosmological information from gravitational lensing. 
Among various statistical methods of gravitational lensing,  
mapping matter distribution on a continuous sky is a key, basic
method 
\citep{2007Natur.445..286M, 2013MNRAS.433.3373V, 2015arXiv150501871C}.
The reconstruction is purely based on deflection of light from sources 
and thus free of conventional astrophysical assumptions.
Non-Gaussian features in the reconstructed
density field are produced
by non-linear gravitational growth and 
can not be extracted by means of the conventional statistics of 
cosmic shear such as two-point correlation function or power spectrum.

Clusters of galaxies serve as a promising probe of cosmology.
The number count of clusters is expected to be highly sensitive to 
growth of matter density perturbations \citep{1992ApJ...386L..33L},
whereas correlation analysis of the position of clusters and cosmic shear
can constrain both the matter density profile and clustering of clusters 
\citep{2012MNRAS.420.3213O,2013ApJ...769L..35O, 2014ApJ...784L..25C}.
The advantage of weak lensing among various techniques is that
it does not rely on physical state of the baryonic component in clusters.
The finding algorithm with cosmic shear analysis is based on 
reconstruction of matter density distribution along a line of sight
\citep{Hamana2004, 2005ApJ...624...59H, 2005AA...442..851M, 2012MNRAS.423.1711M}.
Reconstructed mass density map is utilized to identify high density regions
in the universe that correspond to massive collapsed objects 
such as clusters of galaxies
\citep{2007ApJ...669..714M, 2007AA...462..875S, 2012ApJ...748...56S}.

Previous studies investigated
cosmological information in number counts of weak lensing selected clusters
\citep{2010AA...519A..23M, Kratochvil2010, 
2010MNRAS.402.1049D, Yang2011, 2012MNRAS.426.2870H}.
In our earlier work (\cite{PaperI}, hereafter Paper I),
we examined other statistics 
beyond the abundance of weak lensing selected clusters. 
There, we study the statistical property of weak lensing selected
clusters using realistic mock weak lensing catalogs that incorporate
masked regions and shape noise contaminant.
We developed a theoretical framework based on halo model
that provide robust predictions for the statistical properties of comic shear field
and weak lensing selected clusters.
In this paper, we derive constrain on variants of cosmological models
with the lensing statistics.
We extend our theoretical model to predict the lensing statistics 
for two competing cosmological models (dark energy model and modified gravity model).
In order to realize the realistic situation in galaxy imaging surveys,
we use two hundreds mock weak lensing catalogs 
with the proposed sky coverage of ongoing Subaru Hyper Suprime-Cam (HSC) survey\footnote{
{\rm{http://www.naoj.org/Projects/HSC/index.html}} 
} 
(see Paper I for details).
We derive accurate covariance matrices between the lensing statistics using our large mock catalogs.
We then make realistic forecast for cosmological constraints on 
two competing scenarios with weak lensing selected clusters.

The paper is organized as follows.
In Section~\ref{sec:model}, we briefly describe 
the basics of two cosmological models called 
dark energy model and modified gravity model.
There, we describe in detail the evolution of background and linear density perturbations
in each model.
We also summarize the statistical property of weak lensing selected clusters
and the theoretical model of weak lensing statistics in Section~\ref{sec:WL}.
By using our theoretical model and a large set of mock catalogs,
we make forecast for constraining on cosmological models 
by our method in the upcoming HSC survey in Section~\ref{sec:forecast}.
Conclusions and discussions are summarized in Section~\ref{sec:con}.

\section{COSMOLOGICAL MODEL}\label{sec:model}

There exist various extensions to the standard $\Lambda$CDM model.
We consider two competing models: $w$CDM model and $f(R)$ model.
The former corresponds to a cosmological model based on General Relativity with dark energy
and the latter represents a model with modified gravity.
Either model can explain 
the observed acceleration of cosmic expansion at $z \simlt 1$ 
with appropriate parameters.
Throughout this paper, we assume a spatially flat universe.

\subsection{$w$CDM model}

The expansion of the universe is described by the scale factor $a(t)$.
We adopt the usual normalization $a=1$ at present.
In General Relativity under a Friedmann Robertson Walker (FRW) metric, 
one can derive the time evolution of $a(t)$ as
\beqa
H(a) 
&\equiv& 
\frac{1}{a}\frac{{\rm d}a}{{\rm d}t} \nonumber \\
&=&H_{0}
\Bigg\{
\Omega_{{\rm m}0}a^{-3}+\Omega_{\rm DE}
\nonumber \\
&&
\times
\exp\left[-3\int_{1}^{a} {\rm d}a^{\prime} (1+w_{\rm DE}(a^{\prime}))/a^{\prime}\right]
\Bigg\}^{1/2},
\eeqa
where $H(a)$ is known as the Hubble parameter and 
$H_{0}=100h\, {\rm km}\, {\rm s}^{-1}\, {\rm Mpc}^{-1}$ is the present value of $H(a)$.
The equation of state of dark energy is specified as 
$w_{\rm DE}(a) = P_{\rm DE}(a)/\rho_{\rm DE}(a)$ 
where $P_{\rm DE}$ and $\rho_{\rm DE}$ denote to
pressure and density of dark energy, respectively.

In a $w$CDM model, 
the linear growth rate of matter density fluctuations after matter domination
is given by the solution of the following equation \citep{1998ApJ...508..483W}:
\beqa
\frac{{\rm d}^2 g_{+}}{{\rm d}\ln a^2}
&+&
\left[\frac{5}{2}-\frac{3}{2}w_{\rm DE}(a)\Omega_{\rm DE}(a)\right]\frac{{\rm d} g_{+}}{{\rm d}\ln a}
\nonumber \\
&&\hspace{36pt}
+\frac{3}{2}\left[1-w_{\rm DE}(a)\right]\Omega_{\rm DE}(a)g_{+} = 0, \label{eq:linear_g_wCDM}
\eeqa
where $g_{+}(a) \equiv D(a)/a$ and $D(a)$ represents the linear growth rate of matter density perturbations.
We can obtain $D(a)$ by solving Eq.~(\ref{eq:linear_g_wCDM}) 
with the boundary conditions $g_{+}(a)=1$ and ${\rm d}g_{+}/{\rm d}\ln a=0$ at $a \ll 1$.
The linear density perturbations are commonly characterized by
the linear power spectrum $P_{\rm m}^{L}(k, a)$.
In matter domination, 
the power spectrum of primordial curvature perturbation 
is related to that of the matter density through Poisson equation as follows:
\beqa
\frac{4\pi k^3 P_{\rm m}^{L}(k, a)}{(2\pi)^3} &=&  
\Delta^2_{\cal R}(k_0) \left( \frac{2c^2 k^2}{5H_{0}^{2}\Omega_{\rm m0}} \right)^2 D^2(a) T^2(k) 
\nonumber \\
&&\hspace{60pt}
\times
\left(\frac{k}{k_{0}} \right)^{-1+n_s}, \\
\Delta^2 _{\cal R} (k) &=& \frac{4\pi k^3 P_{\cal R}(k)}{(2\pi)^3},
\eeqa
where $T(k)$ is the transfer function of matter density fluctuations
and  
$P_{\cal R}(k)$ represents power spectrum of curvature perturbation.
In this paper, we normalize $P_{\rm m}^{L}$ with the value of 
$A_{s} \equiv \Delta^2 _{\cal R}(k_{0})$ at $k_{0}=0.002\, {\rm Mpc}^{-1}$ 
as follows in \citet{Hinshaw2013}.

Throughout this paper, 
we consider the simplest model of dark energy 
with constant value of $w_{0}$.
The non-linear gravitational growth of matter density fluctuations
in the model have been investigated by previous numerical studies.
For various $w_{0}$, a large set of cosmological simulations have been used
to derive accurate non-linear power spectrum
\citep{Takahashi2012, 2014ApJ...780..111H}.
Also, \citet{2011ApJ...732..122B} have studied the abundance of dark matter halos 
for a wide range of cosmological parameters in $w$CDM model.
In this work, 
we pay special attention in order to derive observational constraints on
$w_{0}$ from non-linear cosmological information.

\subsection{$f(R)$ model}

In $f(R)$ model, 
the Einstein-Hilbert action is modified by a general function of the scalar curvature $R$,
\beqa
S_{G} = \int {\rm d}^{4}x \sqrt{-g}\left[\frac{R+f(R)}{16\pi G}\right]. \label{eq:action_fR}
\eeqa
The action with Eq.~(\ref{eq:action_fR}) leads the modified Einstein equation as
\beqa
G_{\mu \nu} + f_{R}R_{\mu \nu}-\left(\frac{f}{2}-\Box f_{R} \right)g_{\mu \nu}-\nabla_{\mu}\nabla_{\nu}f_{R} 
\nonumber \\
=8\pi G T_{\mu \nu},
\eeqa
where $f_{R} \equiv {\rm d}f/{\rm d}R$.
Assuming a FRW metric, one can determine the time evolution of the Hubble
parameter in $f(R)$ model 
as follows:
\beqa
H^{2}-f_{R}\left(H\frac{{\rm d}H}{{\rm d}\ln a}+H^2 \right)
+\frac{f}{6}+H^{2}f_{RR}\frac{{\rm d}R}{{\rm d}\ln a} 
\nonumber \\
= \frac{8\pi G}{3}\rho_{m}. \label{eq:H_fR}
\eeqa

One can also consider the evolution of matter density perturbations in $f(R)$ model.
For sub-horizon modes $(k \simgt aH)$ in the quasi-static limit\footnote{
\citet{2008PhRvD..77l3515D} have shown that 
the quasi-static approximation becomes quite reasonable
for models with $|f_{R}| \ll 1$ today. 
}, 
the linear growth of matter density perturbations is
determined by \citep{2007PhRvD..75f4020B}
\beqa
\frac{{\rm d}^2 g_{+}}{{\rm d} a^2}
+\left(\frac{3}{a}+\frac{1}{H}\frac{{\rm d}H}{{\rm d}a}\right)\frac{{\rm d} g_{+}}{{\rm d} a}
-\frac{3\tilde{\Omega}_{\rm m0}a^{-3}}{\left(H/H_0\right)^2 \left(1+f_{R}\right)}
\nonumber \\
\times 
\left(\frac{1-2Q}{2-3Q}\right)\frac{g_{+}}{a^2}= 0, \label{eq:linear_g_fR}
\eeqa
where $\tilde{\Omega}_{\rm m0}$ is the effective matter density at present time.
We can specify this effective density $\tilde{\Omega}_{\rm m0}$ as
\beqa
H_{f(R)} &=& H_{0}\Bigg\{\tilde{\Omega}_{{\rm m}0}a^{-3}+\tilde{\Omega}_{\rm DE}
\nonumber \\ 
&&\times
\exp\left[-3\int_{1}^{a} {\rm d}a^{\prime} (1+\tilde{w}_{\rm DE}(a^{\prime}))/a^{\prime}\right]\Bigg\}^{1/2},
\eeqa
where $H_{f(R)}$ is given by Eq.~(\ref{eq:H_fR}).
The function $Q$ in Eq.~(\ref{eq:linear_g_fR}) is given by
\beqa
Q(k, a) = -2\left(\frac{k}{a}\right)^2 \frac{f_{RR}}{1+f_{R}}.
\eeqa
Note that the function of $Q$ induces non-trivial scale
dependence of the linear growth rate $g_{+}(k, a)=D(k, a)/a$ in $f(R)$ model,
while the linear growth rate is solely a function of $a$ in General Relativity.

In this paper, we will consider the representative example of $f(R)$ model
proposed by \citet{2007PhRvD..76f4004H} (hereafter denoted as HS model),
\beqa
f(R) = -2\Lambda \frac{R^{n}}{R^{n}+\mu^{2n}},
\eeqa
where $\Lambda$, $\mu$ and $n$ are free parameters in the model.
For $R\gg \mu^2$, 
one can approximate the function of $f(R)$ as follows:
\beqa
f(R) = -2\Lambda -\frac{f_{R0}}{n}\frac{\bar{R}_{0}^{n+1}}{R^n},
\eeqa
where $\bar{R}_{0}$ is defined 
by the present scalar curvature of the background space-time
and $f_{R0} = -2\Lambda \mu^2/\bar{R}_{0}^2 = f_{R}(\bar{R}_{0})$.
In the HS model with $|f_{R0}| \ll 1$, 
the background expansion behaves similarly to the standard $\Lambda$CDM model.
In practice, for $|f_{R0}| \ll 10^{-2}$, 
geometric tests such as measurement of supernovas 
cannot distinguish with HS model and the $\Lambda$CDM model \citep{2012PhRvD..85b4006M}.
Nevertheless, measurement of gravitational growth
would be helpful to constrain on HS model if
the scale dependence of growth rate as shown in Eq.~(\ref{eq:linear_g_fR})
is detected. 
The non-linear gravitational growth in HS model have been studied with cosmological simulations
\citep{2008PhRvD..78l3524O, 2009PhRvD..79h3518S, 2013PhRvD..88j3507H, 2014ApJS..211...23Z}.
These studies suggest that 
statistics of galaxy groups or clusters 
provide meaningful information about the modification of gravity.
Therefore, a combination of statistics of cosmic shear and galaxy clusters
can offer an interesting probe of HS model.
In the following, we focus on the case of $n=1$ 
for which many numerical studies have been performed.
 
\section{Weak lensing}\label{sec:WL}
Here, we summarize basics of weak gravitational lensing 
and the finding algorithm of galaxy clusters
with weak lensing measurement.
Further details of the statistical properties of weak lensing selected clusters
are found in Paper I.

\subsection{Basics}

When considering the observed position of a source object as $\mbox{\boldmath $\theta$}$ 
and the true position as $\mbox{\boldmath $\beta$}$,
one can express the distortion of image of a source object by the following 2D matrix:
\beqa
A_{ij} = \frac{\partial \beta^{i}}{\partial \theta^{j}}
           \equiv \left(
\begin{array}{cc}
1-\kappa -\gamma_{1} & -\gamma_{2}  \\
-\gamma_{2} & 1-\kappa+\gamma_{1} \\
\end{array}
\right), \label{distortion_tensor}
\eeqa
where $\kappa$ is convergence and $\gamma$ is shear.

Let us consider the case of General Relativity. 
In this case, one can relate each component of $A_{ij}$ to
the second derivative of the gravitational potential as follows
\citep{Bartelmann2001, Munshi2008};
\beqa
A_{ij} &=& \delta_{ij} - \Phi_{ij}, \label{eq:Aij} \\
\Phi_{ij}  &=&\frac{2}{c^2}\int _{0}^{\chi}{\rm d}\chi^{\prime} g(\chi,\chi^{\prime}) \partial_{i}\partial_{j}\Phi(\chi^{\prime}), \label{eq:shear_ten}\\	
g(\chi,\chi^{\prime}) &=& \frac{r(\chi-\chi^{\prime})r(\chi^{\prime})}{r(\chi)},
\eeqa
where $\chi$ is the comoving distance and $r(\chi)$ represents the comoving angular diameter distance.
Gravitational potential $\Phi$ can be related to matter density perturbation
$\delta$ according to Poisson equation.
Therefore, convergence can be expressed as the weighted integral
of $\delta$ along the line of sight;
\beqa
\kappa = \frac{3}{2}\left(\frac{H_{0}}{c}\right)^2 \Omega_{\rm m0} \int _{0}^{\chi}{\rm d}\chi^{\prime} g(\chi,\chi^{\prime}) \frac{\delta}{a}. \label{eq:kappa_delta}
\eeqa

Next, we consider weak gravitational lensing effect in HS model.
When adopting the Newtonian gauge, one can express the line element as
\beqa
{\rm d}s^2 = a(\eta)^2 \left[ -\left(1+2\Phi\right){\rm d}\eta^2+\left(1-2\Psi\right){\rm d}{\bd x}^2\right],
\eeqa
where $\eta$ is the conformal time.
In this metric, the deflection angle by gravitational lensing can be written as \citep{Bartelmann2001, Munshi2008},
\beqa
\bd \alpha = \frac{2}{c^2} \int \nabla_{\perp}\left(\frac{\Phi+\Psi}{2}\right){\rm d}\ell,
\eeqa
where $\nabla_{\perp}$ corresponds to the perpendicular component of the gradient along a line of sight 
for proper distance and ${\rm d}\ell$ represents the line integral in terms of proper distance.
Considering the modified Einstein equation in HS model with $|\Phi| \ll 1$, $|\Psi| \ll 1$, and $|f_{R0}| \ll 1$,
one can derive the following equations of $\Phi$ and $\Psi$
(the derivation is found in e.g., \citet{2014MNRAS.440..833A});
\beqa
\frac{1}{a^2}\nabla^2 \Phi &=& \frac{16\pi G}{3}\delta \rho-\frac{1}{6}\delta R, \\
\frac{1}{a^2}\nabla^2 \Psi &=& \frac{8 \pi G}{3}\delta \rho+\frac{1}{6}\delta R,
\eeqa
where $\nabla$ is the gradient with respect to comoving distance 
and $\delta\rho$ and $\delta R$ 
represent the density perturbation 
and the fluctuation of the scalar curvature $R$, respectively.
Hence, the lensing potential $(\Phi+\Psi)/2$ follows the same equation in General Relativity as
\beqa
\frac{1}{a^2}\nabla^2 \frac{\Phi+\Psi}{2} = 4\pi G \delta\rho.
\eeqa
These results show that 
Eqs.~(\ref{eq:Aij}), (\ref{eq:shear_ten}) and 
(\ref{eq:kappa_delta}) are available 
in the HS model with $|f_{R0}| \ll 1$.

\subsection{Cluster selection}
\label{subsec:WL_select_cluster}
Weak lensing is a powerful tool to reconstruct the projected matter density field.
The conventional technique for reconstruction is 
based on the smoothed map of cosmic shear.
Let us first define the smoothed convergence field as
\beqa
{\cal K} (\bd{\theta}) = \int {\rm d}^2 \phi \ \kappa(\bd{\theta}-\bd{\phi}) U(\bd{\phi}), \label{eq:ksm_u}
\eeqa
where $U$ is the filter function to be specified below.
We adopt the compensated Gaussian filter for $U$ as
\beqa
U(\theta) = \frac{1}{\pi \theta_{G}^2}
\exp\left(-\frac{\theta^2}{\theta_{G}^2}\right)
-\frac{1}{\pi \theta_{o}^2}
\left[1-\exp\left(-\frac{\theta_{o}^2}{\theta_{G}^2}\right)\right],
\label{eq:compensated_Gauss}
\eeqa
where $\theta_{o}$ represents the boundary of the filter
and we set $U$ to be zero for $\theta > \theta_{o}$.

For a given function of $U$,
the power spectrum of a noise convergence field ${\cal N}$ 
is given by \citep{VanWaerbeke2000}
\beqa
P_{\cal N}(\ell) = \frac{\sigma_{\gamma}^2}{2n_{\rm gal}}
|{\tilde U}(\ell)|^2
\label{eq:noise_power}
\eeqa
where $\sigma_{\gamma}$ is the rms of the intrinsic source ellipticities
$n_{\rm gal}$ represents the number density of source galaxies,
and ${\tilde U}$ is the Fourier transform of $U$.
From Eq.~(\ref{eq:noise_power}), we define the moment of ${\cal N}$ as
\beqa
\sigma_{{\rm noise}, i} = 
\left(\int \frac{{\rm d}^2 \bd{\ell}}{(2\pi)^2}\, \ell^{2i} P_{\cal N}(\ell)\right)^{1/2}.
\label{eq:noise_moment}
\eeqa
Throughout this paper, 
we set $\sigma_{\gamma}=0.4$, $n_{\rm gal}=10\, {\rm arcmin}^{-2}$
and assume the source redshift of $z_{\rm source}=1$.
These are typical values for ground-based galaxy imaging surveys
\citep{2012MNRAS.427..146H}. 
Also, we adopt 
the smoothing scale of $\theta_{G}=5/\sqrt{8\ln 2}=2.12$ arcmin
and $\theta_{o}=30$ arcmin.
This leads to $\sigma_{\rm noise, 0}\simeq 0.017$.
The set up is examined in detail in Paper I with numerical simulations.

On a smoothed lensing map, 
convergence peaks with high signal-to-noise ratio $\nu = {\cal K}/\sigma_{{\rm noise}, 0}$ 
are likely caused by galaxy clusters
\citep{Hamana2004}.
We thus locate high-$\nu$ peaks on a ${\cal K}$ map 
and associate each of them with an isolated massive halo along the line of sight.
For a dark matter halo, we assume the universal NFW density profile \citep{Navarro1997}.
We adopt the functional form of the concentration parameter 
in \citet{2008MNRAS.390L..64D},
\beqa
c_{\rm vir}(M, z) = 5.72 \left( \frac{M}{10^{14} h^{-1}M_{\odot}}\right)^{-0.081}(1+z)^{-0.71}.
\eeqa
Note that the NFW density profile 
provides a reasonable fit also for halos in HS model with $|f_{R0}| \ll 1$
\citep{2009PhRvD..79h3518S,2012PhRvD..85l4054L}
and that the corresponding convergence $\kappa_{h}$ can be calculated analytically 
\citep{Hamana2004}.

In order to predict peak heights in a ${\cal K}$ map, 
we adopt the simple assumption that each peak position is exactly at the halo center.
Under this assumption, the peak height in absence of shape noise is given by
\beqa
{\cal K}_{{\rm peak}, h} = \alpha \int {\rm d}^2\phi \, U(\phi; \theta_{G},\theta_{o})\kappa_{h}(\phi) + \beta, 
\label{eq:kpeak_nfw}
\eeqa
where $\alpha=0.9$ and $\beta=0$ are found to be in good agreement with numerical simulations as shown in Paper I.
The actual peak height on a noisy ${\cal K}$ map 
is determined not by Eq.~(\ref{eq:kpeak_nfw}), 
but by a probability distribution function \citep{2010ApJ...719.1408F}.
The probability distribution function of the measured peak height ${\cal K}_{\rm peak, obs}$
with a given ${\cal K}_{{\rm peak}, h}$ is denoted
by ${\rm Prob}({\cal K}_{\rm peak, obs}|{\cal K}_{{\rm peak}, h})$ in this paper.
The detailed functional form of ${\rm Prob}({\cal K}_{\rm peak, obs}|{\cal K}_{{\rm peak}, h})$
is found in Paper I.
In Paper I, we show that our model of ${\rm Prob}({\cal K}_{\rm peak, obs}|{\cal K}_{{\rm peak}, h})$
is quite reasonable for peaks with high signal-to-noise ratio
in the case of $\theta_{G}\sim 2\, {\rm arcmin}$, 
$\sigma_{\gamma}=0.4$, 
and $n_{\rm gal} \simgt 10\, {\rm arcmin}^{-2}$.

\subsection{Statistics}
\label{subsec:WL_stat}
We here summarize a set of statistics obtained from weak lensing measurement.
Table \ref{tab:model_summary} shows the summary of our model of each lensing statistic.

\begin{table*}
\begin{center}
\begin{tabular}{|l|l|l|l|}
\hline
Statistic & Definition & Integrand in halo model & Reference \\ \hline
Convergence power spectrum $P_{\kappa\kappa}$ & Eq.~(\ref{eq:kappa_power}) & Matter power spectrum $P_{\delta}$ & \citet{Takahashi2012} ($w$CDM) \\
& & & \citet{2014ApJS..211...23Z} ($f(R)$)  \\ \hline
Convergence peak count $N_{\rm peak}$ & Eq.~(\ref{eq:npeak}) & Halo mass function ${\rm d}n/{\rm d}M$ &\citet{2011ApJ...732..122B} ($w$CDM) \\
& & & \citet{2011PhRvD..84h4033L} ($f(R)$)  \\ \hline
Convergence peak auto spectrum $P_{\rm pp}$ & Eq.~(\ref{eq:peak_power}) & Halo mass function ${\rm d}n/{\rm d}M$ & \citet{2011ApJ...732..122B} ($w$CDM) \\
& & Linear halo bias $b_h$ & \citet{2011PhRvD..84h4033L} ($f(R)$)  \\ \hline
Convergence peak cross spectrum $P_{{\rm p}\kappa}$ & Eqs.~(\ref{eq:peak_kappa_power1}) and (\ref{eq:peak_kappa_power2})
& Halo mass function ${\rm d}n/{\rm d}M$ & \citet{2011ApJ...732..122B} ($w$CDM) \\
& & Linear halo bias $b_h$ & \citet{2011PhRvD..84h4033L} ($f(R)$)  \\
\hline
\end{tabular}
\caption{
  We summarize the elements in our model to derive
  the lensing statistics presented in this paper.
  Each column shows, the statistical quantity of interest,
  the definition, the integrand in our model
  and the reference of fitting formula to compute the integrand.
}
\label{tab:model_summary}
\end{center}
\end{table*}

\subsubsection{Convergence power spectrum}

First, we consider convergence power spectrum,
which is a direct probe of the underlying matter density field.
Under the Limber approximation\footnote{
The validity of Limber approximation have been discussed in e.g., \citet{2009PhRvD..80l3527J}.
The typical accuracy of Limber approximaion is of a level of $\simlt$1\% for $\ell > 10$.
} \citep{Limber:1954zz,Kaiser:1991qi}
and Eq.~(\ref{eq:kappa_delta}), 
one can calculate the convergence power spectrum as 
\beqa
P_{\kappa\kappa}(\ell) &=& \int_{0}^{\chi_s} {\rm d}\chi \frac{W_{\kappa}(\chi)^2}{r(\chi)^2} 
P_{\delta}\left(k=\frac{\ell}{r(\chi)},z(\chi)\right)
\label{eq:kappa_power},
\eeqa
where $P_{\delta}(k)$ is the three dimensional matter power spectrum, 
$\chi_s$ is comoving distance to source galaxies and 
$W_{\kappa}(\chi)$ is the lensing weight function defined as
\beqa
W_{\kappa}(\chi) = \frac{3}{2}\left(\frac{H_{0}}{c}\right)^2 \Omega_{\rm m0}
\frac{r(\chi_s-\chi)r(\chi)}{r(\chi_s)}(1+z(\chi)).
\eeqa

The non-linear gravitational growth of $P_{\delta}(k)$ significantly affects 
the amplitude of convergence power spectrum at the angular scales less than 1 degree
\citep{2000ApJ...530..547J, Hilbert2009, Sato2009}.
Therefore, accurate theoretical prediction of the non-linear matter power spectrum 
is essential for cosmological constraints from weak lensing power spectrum.
In order to predict the non-linear evolution of $P_{\delta}(k)$ 
for the standard $\Lambda$CDM universe,
numerical approach based on $N$-body simulations has 
been employed extensively over the past few decades
\citep{1996MNRAS.280L..19P, 2003MNRAS.341.1311S, 2010ApJ...715..104H, Takahashi2012}.
In particular, \citet{Takahashi2012} provides an accurate fitting formula
of non-linear $P_{\delta}(k)$ for various $w$CDM models.
We adopt their model in the following.
For the HS model of modified gravity,
\citet{2008PhRvD..78l3524O, 2013PhRvD..88j3507H, 2014ApJS..211...23Z} 
investigated the non-linear evolution of $P_{\delta}(k)$.
Recently, \citet{2014ApJS..211...23Z} derived
a new fitting formula of non-linear $P_{\delta}(k)$ 
with a large set of numerical simulations.
We use the model in \citet{2014ApJS..211...23Z} for the HS model.

\subsubsection{Convergence peak count}

We identify local maxima in a smoothed lensing map and 
match each peak with a massive dark matter halo along the line of sight.
The simple peak count is useful to extract the information of the abundance
of massive clusters. 
One can select the lensing peaks by its peak height.
We define the peak signal-to-noise ratio 
by $\nu = {\cal K}_{\rm peak, obs}/\sigma_{\rm noise,0}$.
For a given threshold $\nu_{\rm thre}$, 
one can predict the surface number density of peaks
with $\nu > \nu_{\rm thre}$ as follows
\citep{Hamana2004}:
\beqa
N_{\rm peak}(\nu_{\rm thre}) 
=
\int {\rm d}z\, {\rm d}M\, 
\frac{{\rm d}^2V}{{\rm d}z{\rm d}\Omega} \frac{{\rm d}n}{{\rm d} M}\, S(z, M|\nu_{\rm thre}), \label{eq:npeak}
\eeqa
where ${\rm d}n/{\rm d}M$ represents the mass function of dark matter halo 
and the volume element is expressed as ${\rm d}^2 V/{\rm d}z{\rm d}\Omega = \chi^2/H(z)$
for a spatially flat universe.
In Eq.~(\ref{eq:npeak}), $S(z, M|\nu_{\rm thre})$ expresses the selection function
of weak lensing selected clusters
for a given threshold of $\nu_{\rm thre}$. It is given by
\beqa
S(z, M|\nu_{\rm thre}) 
&=& 
\int_{\nu_{\rm thre}\sigma_{\rm noise,0}}^{\infty} 
{\rm d}{\cal K}_{\rm peak, obs}\, 
\nonumber \\
&&\times 
{\rm Prob}({\cal K}_{\rm peak, obs}|\, {\cal K}_{{\rm peak},h}(z, M)).
\eeqa

Throughout this paper, we adopt the model of halo mass function in \citet{2011ApJ...732..122B} 
for $w$CDM models and the model in \citet{2011PhRvD..84h4033L} for HS models.

\subsubsection{Convergence peak auto spectrum and cross spectrum}

Next, we consider the auto-correlation function of peaks and 
the cross correlation of peaks and convergence.
We expect that these statistics contain information of dark matter density profile around
clusters as well as of the clustering of clusters.

In Paper I, 
by using halo model approach,
we derive the auto power spectrum of weak lensing selected clusters 
for a given threshold $\nu_{\rm thre}$ as follows:
\beqa
P_{\rm pp}(\ell) = \int {\rm d}\chi\, \frac{W_{\rm p}(\chi|\nu_{\rm thre})^2}{r(\chi)^2} 
P^{L}_{m}\left(k=\frac{\ell}{r(\chi)}, z(\chi)\right), \label{eq:peak_power}
\eeqa
where the window function $W_{\rm p}(\chi)$ is given by
\beqa
W_{\rm p}(\chi|\nu_{\rm thre}) 
&=& 
\frac{1}{N_{\rm peak}(\nu_{\rm thre})} 
\int {\rm d}M\, 
\frac{{\rm d}^2V}{{\rm d}\chi{\rm d}\Omega} \frac{{\rm d}n}{{\rm d} M} 
\nonumber \\
&& \hspace{36pt} \times 
\, S(z, M|\nu_{\rm thre}) b_{h}(z, M),
\eeqa
where $b_{h}$ is the linear halo bias.
Similarly, the cross power spectrum between peaks and convergence is given by 
(also, see \citet{2011PhRvD..83b3008O})
\beqa
P_{{\rm p}\kappa}(\ell) &=& P^{1h}_{{\rm p}\kappa}(\ell)+P^{2h}_{{\rm p}\kappa}(\ell), \\
P^{1h}_{{\rm p}\kappa}(\ell) &=& 
\frac{1}{N_{\rm peak}(\nu_{\rm thre})} 
\int {\rm d}z\, 
\frac{{\rm d}^2V}{{\rm d}z{\rm d}\Omega} 
\nonumber \\
&& \times
\int {\rm d}M\, \frac{{\rm d}n}{{\rm d}M} S(z, M|\nu_{\rm thre})
\frac{1}{\bar{\rho}_{m}(z)}\frac{W_{\kappa}(\chi)}{r(\chi)^2}
\nonumber \\ 
&& 
\hspace{60pt}
\times  
M \tilde{u}_{m}\left(k=\frac{\ell}{r(\chi)}\Bigg|z(\chi), M\right), \label{eq:peak_kappa_power1} \\
P^{2h}_{{\rm p}\kappa}(\ell) &=&
\int {\rm d}\chi\, \frac{W_{\rm p}(\chi|\nu_{\rm thre})W_{\kappa}(\chi)}{r(\chi)^2}
\nonumber \\ 
&& \hspace{60pt} \times 
P^{L}_{m}\left(k=\frac{\ell}{r(\chi)}, z(\chi)\right), \label{eq:peak_kappa_power2}
\eeqa
where $M \tilde{u}_{m}$ is the Fourier transform of NFW profile.

In this paper, we adopt the functional form of $b_{h}$ proposed
by \citet{2011ApJ...732..122B} for $w$CDM  model.
For HS models, we derive the linear halo bias based on the peak-background
split formalism \citep{Sheth1999}.
For this purpose, we use the model of mass function shown in \citet{2011PhRvD..84h4033L}
and relate the linear halo bias with mass function as follows:
\beqa
b_{h}(z, M) = 1-\frac{\partial \ln f_{\rm MF}}{\partial \delta_{c}},
\eeqa
where $\delta_{c}$ is the linear critical density for spherical collapse model.
$f_{\rm MF}$ is defined by
\beqa
\frac{{\rm d}n}{{\rm d}M}(z, M) = f_{\rm MF}(z, M) \frac{\bar{\rho}_{m}}{M}
\Biggl|\frac{\partial \ln \sigma_{M}^{-1}}{\partial \ln M}\Biggr|,
\eeqa
where $\sigma_{M}$ is the mass variance for HS models.
The functional form of $f_{\rm MF}$ and $\sigma_{M}$ are found in \citet{2011PhRvD..84h4033L}.

\begin{figure*}
\centering \FigureFile(70mm,50mm){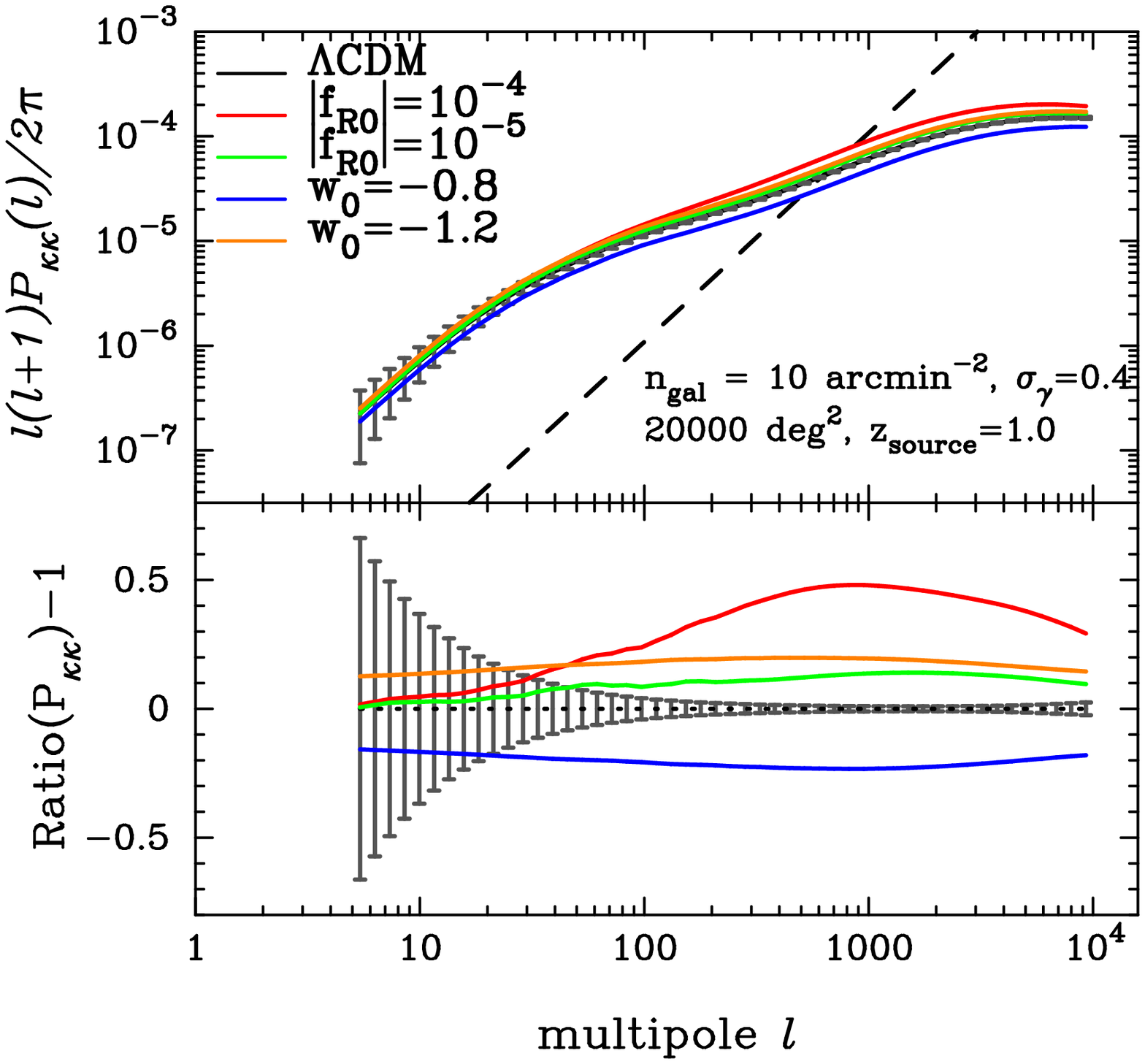}
\centering \FigureFile(70mm,50mm){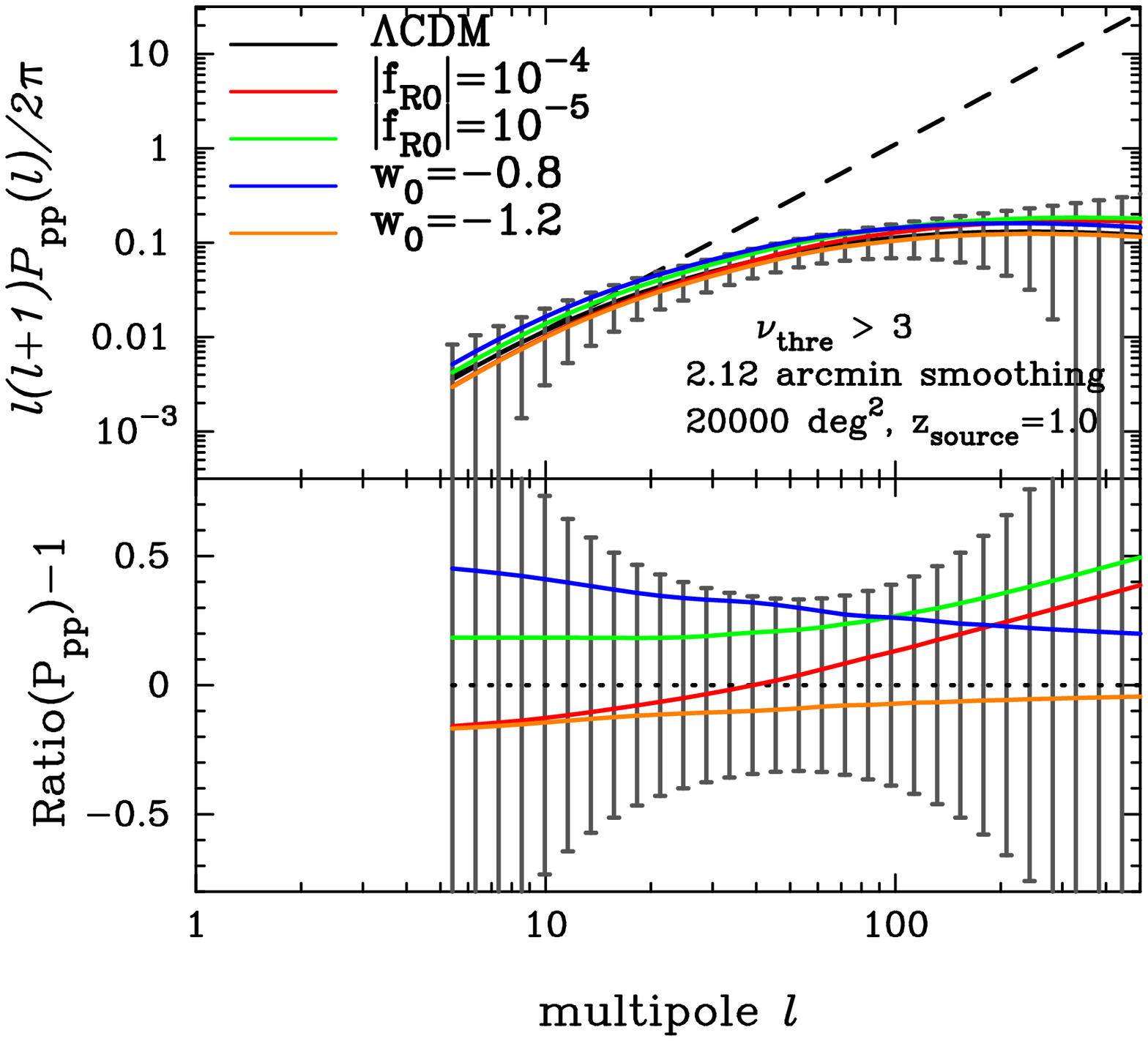}
\centering \FigureFile(70mm,50mm){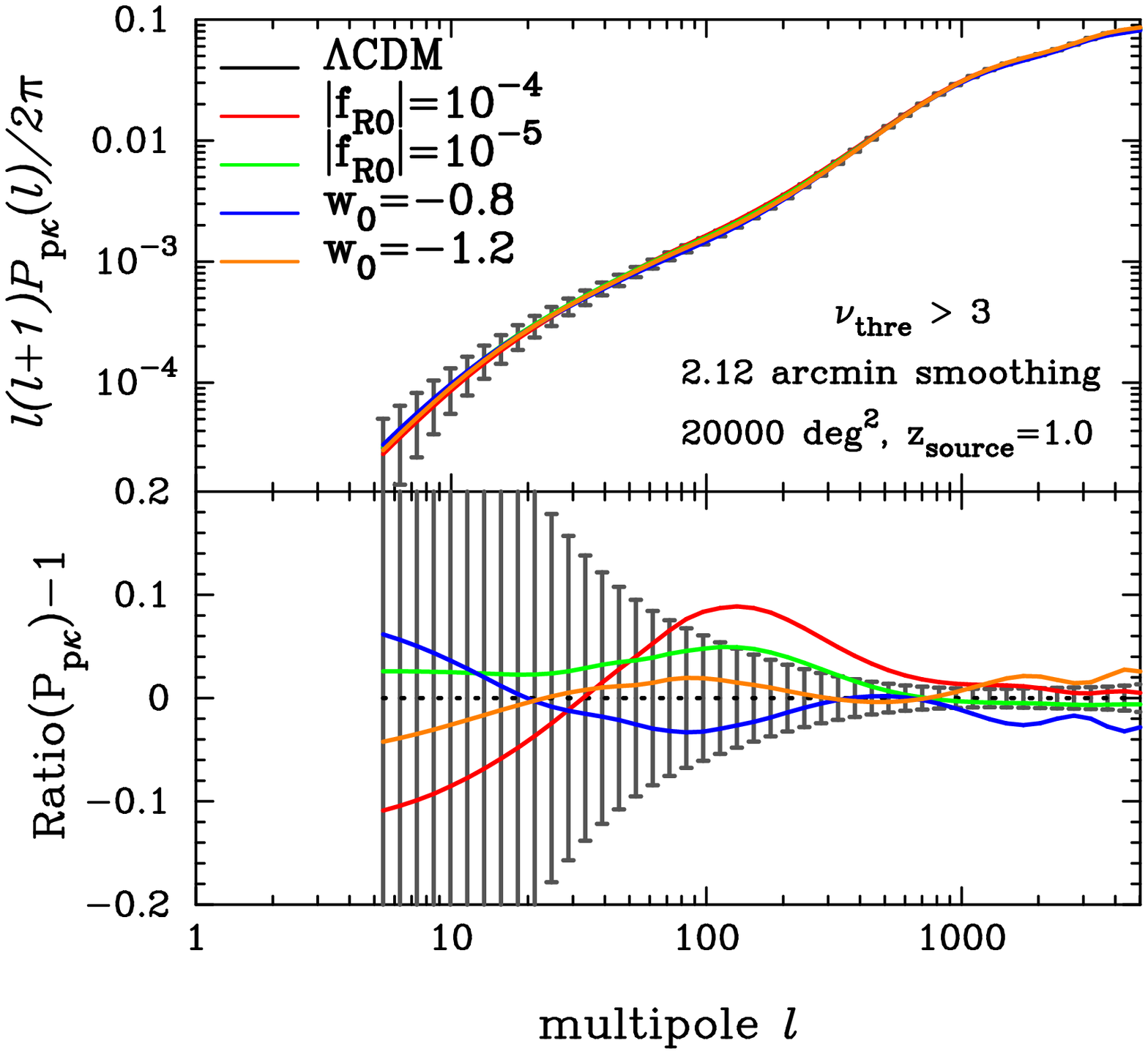}
\centering \FigureFile(70mm,50mm){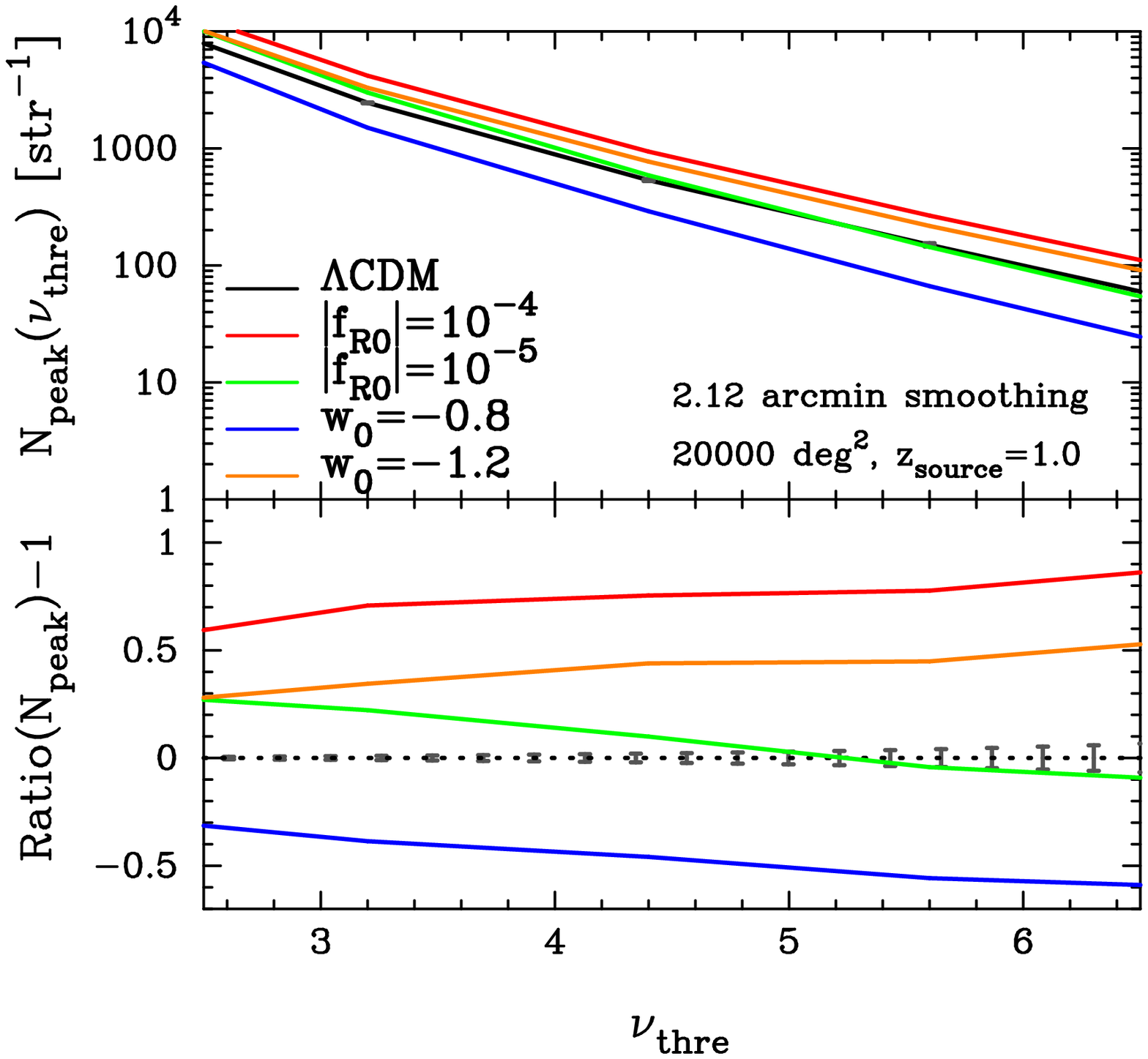}
\caption{
	The dependence of  weak lensing statistics on model parameters.
	We plot the power spectra of convergence (top-left), 
	the power spectra of lensing peaks (top-right),
	the cross spectra between convergence and lensing peaks (bottom-left), 
	and the number counts of lensing peaks (bottom-left).
	We define the lensing peaks as those with signal-to-noise ratio 
	$\nu_{\rm thre} > 3$ in the top right and the bottom left panel.
	The error bars in each panel represent the Gaussian or Poission errors
        in the case of 
	sky coverage of 20,000 square degrees.
	The dashed line in two top panels corresponds to 
	the shot noise term of auto power spectrum.
	In each panel, we show the theoretical model of weak lensing statistics 
	for cosmological models with different value of $w_{0}$ and $|f_{R0}|$.
	The bottom portion in each panel provides the ratio of each statistic between 
	$\Lambda$CDM model and $w$CDM model or $f(R)$ model.
	\label{fig:stat_comp_model}
	}
\end{figure*}

\subsection{Dependence of cosmological model}

Figure \ref{fig:stat_comp_model} summarizes
the dependence of the weak lensing statistics on cosmological parameters.
In order to calculate the lensing statistics for different cosmological models,
we use the description summarized in Section~\ref{subsec:WL_stat} and 
Table~\ref{tab:model_summary}.
The four panels in Figure \ref{fig:stat_comp_model} show $P_{\kappa\kappa}$, $P_{\rm pp}$, $P_{{\rm p}\kappa}$
and $N_{\rm peak}$ for various cosmological models.
In each panel, the black line shows the model prediction for $\Lambda$CDM model, 
whereas the colored line represents the model for $w$CDM models or $f(R)$ models.
In this figure, we adopt the value of $w_{0} = -0.8$ and $-1.2$ 
or $|f_{R0}|=10^{-4}$ and $10^{-5}$.
Other cosmological parameters are consistent with those derived by
WMAP nine-year analysis \citep{Hinshaw2013}.
We also present the ratio of each statistic with respect to the $\Lambda$CDM model,
i.e., we plot
\beqa
{\rm Ratio}(X) \equiv \frac{X(w{\rm CDM\, or\, HS\, model})}{X_{\Lambda{\rm CDM}}},
\eeqa
where $X = P_{\kappa\kappa}, P_{\rm pp}, P_{{\rm p}\kappa}, N_{\rm peak}$.
In Figure \ref{fig:stat_comp_model}, 
we simply consider the Gaussian covariance of power spectra 
and the Poisson error of the number count of lensing peaks
assuming a hypothetical survey with sky coverage of 20,000 squared degrees.

The $w$CDM model mainly affects 
the linear growth rate of matter density perturbations and the cosmic expansion rate.
Thus, it modifies $P_{\kappa\kappa}$ in an almost scale-independent way
and causes larger impact on number of more massive halos, i.e. $N_{\rm peak}$ for higher peaks.

On the other hand, 
the $f(R)$ model induces the scale-dependence of the linear gravitational growth
and reduces gravitational force enhancement as a function of the depth 
of the gravitational potential wells of dark matter halos.
The scale-dependence of growth of linear density fluctuations is determined by
the mass of additional scalar degree of freedom $f_{R}$ in this model.
Essentially, the $f_{R}$ field characterizes two physical length scales:
One is the region where the linear growth rate is quite similar to that of $\Lambda$CDM
and the other is the region where the fluctuation growth is amplified
by the additional fifth force.
The transition between the two would occur at the compton wavelength of $f_{R}$.
When density fluctuations grow, another important 
mechanism, the so-called chameleon mechanism, operates
and induces the density dependence of the mass of $f_{R}$ field.
Even at the small length scales where the linear growth rate is amplified,
the chameleon mechanism can effectively switch the gravitational force 
so that $\Lambda$CDM-like evolution is recovered.
This additional effect would make matter power spectrum and halo mass
function more complex.
Especially, specific features like bumps are found in the matter power
spectrum and in halo mass function on $f(R)$ model
\citep{2011PhRvD..84h4033L,2014ApJS..211...23Z}.
Therefore, the $f(R)$ model amplifies $P_{\kappa\kappa}$ 
in a wavenumber dependent manner,
and changes appreciably the number of halos
with masses near the relevant mass scale of  
the force enhancement.

Although the error estimate 
in Figure \ref{fig:stat_comp_model} might seem optimistic,
$P_{\kappa\kappa}$ and $N_{\rm peak}$ are expected to be useful to 
extract cosmological information of the deviation from $\Lambda$CDM model.
Note, however, that the statistics also depend on parameters
other than $w_{0}$ and $f_{\rm R0}$.
For example, the convergence power spectrum has
strong degeneracy among cosmological parameters, 
while the number of lensing peaks can be much dependent of
the concentration of dark matter density profile.
In order to overcome the degeneracies and uncertainties, 
we utilize the cross power spectrum of convergence and
lensing peaks $P_{\rm p \kappa}$\footnote{
In principle, the auto power spectrum of lensing peaks $P_{\rm pp}$
can be a direct measure of clustering of clusters.
However, we find that the signal to noise ratio of $P_{\rm pp}$ 
is significantly smaller than that of $P_{{\rm p}\kappa}$.
In the case of lensing surveys with 
the sky coverage of $\sim$1000 squared degrees,
we can find only $\sim100$ peaks at most.
Thus, the auto-power spectrum $P_{\rm pp}$ would be dominated by the Poisson noise.
We expect that $P_{\rm pp}$ can be useful to extract some information 
only if the survey area exceeds $\sim$10,000 squared degrees
}.
The cross correlation can be used to distinguish two different information contents,
i.e., cosmological parameters and the dark matter density profile.
The former can be mainly extracted from the cross correlation at degree scales
whereas the latter can be derived from the cross correlation signals at arcmin scales.
Therefore, combined analysis with 
$P_{\kappa\kappa}, N_{\rm peak}$ and $P_{\rm p \kappa}$
is a powerful tool of cosmology with weak lensing selected clusters.

\section{FORECAST}\label{sec:forecast}

We perform a Fisher analysis to make forecast for 
constraints on $w$CDM model or $f(R)$ model
with the ongoing Hyper Suprime-Cam (HSC) survey.

Let us briefly summarize the Fisher analysis.
For a multivariate Gaussian likelihood, 
the Fisher matrix $F_{ij}$ can be written as
\beqa
F_{ij} = \frac{1}{2} {\rm Tr} \left[ A_{i} A_{j} + C^{-1} H_{ij} \right], \label{eq:Fij}
\eeqa
where $A_{i} = C^{-1} \partial C/\partial p_{i}$, 
$H_{ij} = 2 \left(\partial \mu/\partial p_{i} \right)\left(\partial \mu/\partial p_{j} \right)$, 
$C$ is the data covariance matrix, 
$\mu$ represents the assumed model, 
and 
$\bd{p}$ describes parameters of interest.
In the present study, 
we consider only the second term in Eq.~(\ref{eq:Fij}).
Because $C$ is expected to scale proportionally inverse to the survey area, 
the second term will be dominant 
for a large area survey (see, e.g., \citet{2009AA...502..721E}). 
We consider the following set of parameters to vary:
$\bd{p} = (10^{9}A_{s}, n_{s}, \Omega_{\rm m0}, A_{\rm vir, 0},
w_{0}\, {\rm or}\, |f_{\rm R0}|)$
where $A_{\rm vir, 0}$ represents the normalization of concentration
of dark matter density profile.
We define $A_{\rm vir, 0}$ as 
\beqa
c_{\rm vir} = A_{\rm vir, 0} \left( \frac{M}{10^{14} h^{-1}M_{\odot}}\right)^{-0.081}(1+z)^{-0.71}.
\eeqa

We construct the data vector ${\bd D}$ from a set of 
binned spectra $P_{\kappa \kappa}, P_{{\rm p}\kappa}$ and 
the number count of peaks $N_{\rm peak}$ as, 
\beqa
{D_{i}} = \{P_{\kappa\kappa}(\ell_{1}),...,P_{\kappa\kappa}(\ell_{10}),
P_{{\rm p}\kappa}(\ell_{1}),...,P_{{\rm p}\kappa}(\ell_{10}), 
\nonumber \\
N_{\rm peak}(\nu_{{\rm thre},1}),...,N_{\rm peak}(\nu_{{\rm thre}, 6})\}, \label{eq:Dvec}
\eeqa
where $P_{{\rm p}\kappa}$ is defined by the cross spectrum between convergence
and lensing peaks with $\nu_{\rm thre} = 3$.
For the Fisher analysis, we use 10 bins in the range of $\ell_{i} = [20,2000]$
and 6 bins in the range of $\nu_{{\rm thre},i} = [2.5, 5.5]$.
In total, a data vector has 26 elements, 
$2\times10$ for power- and cross-spectra and 6 for peak counts.
In the Fisher analysis, we need the derivative of statistics
$D_{i}$ with respect to $\bd{p}$.
We evaluate the numerical derivatives as follows:
\beqa
\frac{\partial D_{i}}{\partial p_{a}} 
= \frac{D_{i}(p^{(0)}_{a}+dp_{a})-D_{i}(p^{(0)}_{a}-dp_{a})}{2dp_{a}},
\label{eq:dev}
\eeqa
where $p^{(0)}_{a}$ is the fiducial value, $dp_{a}$ is the variation of $a$-th parameter
and we use the similar definition to $P_{{\rm p}\kappa}$ and $N_{\rm peak}$.
In order to evaluate the term $\partial D_{i}/\partial p_{a}$ in Eq.~(\ref{eq:dev}),
we calculate the relevant statistics adopting the analytic models shown in 
Section~\ref{subsec:WL_stat} and Table~\ref{tab:model_summary} for different cosmologies.
We summarize the fiducial value of $\bd{p}$ and $d\bd{p}$ in Table \ref{tab:param}.

\begin{table}
\begin{center}
\begin{tabular}{|c|c|c|c|c|c|c|}
\hline
 & $10^{9}A_{s}$ & $n_{s}$ & $\Omega_{\rm m0}$ & $A_{\rm vir, 0}$ & $w_{0}$ & $10^{5}|f_{\rm R0}|$ \\ \hline
fiducial & 2.41 & 0.972 & 0.279 & 5.72 & -1 & 1 \\ \hline
$d\bd{p}$ & 0.1 & 0.13 & 0.01 & 1.0 & 0.2 & 0.5 \\ \hline
\end{tabular}
\caption{
Parameters for Fisher analysis.
We denote by $d\bd{p}$ the variation of parameters around the fiducial vaule
that is used for estimation of the numerical derivative of 
$P_{\kappa \kappa}$, $P_{{\rm p}\kappa}$ and $N_{\rm peak}$.
	}
\label{tab:param}
\end{center}
\end{table}

We need a $26 \times 26$ data covariance matrix 
for the Fisher analysis.
In order to derive the covariances, 
we use 200 masked sky simulations genearated in Paper I.

The masked sky simulations are constructed 
based on ten independent full-sky weak lensing simulations.
We perform the full-sky weak lensing simulations
combined with a large set of dark matter $N$-body simulations
with consistent cosmological model with WMAP nine-year results \citep{Hinshaw2013}.
In each simulation, we properly take into account the deflection 
of light path on a curved sky.
From each full-sky realization,
we extract multiple ``observed" regions with similar survey
geometry to the Hyper-Suprime Cam (HSC) survey.
We create 20 convergence maps on masked HSC sky by
selecting the desired sky coverage (565 + 680 = 1245 squared degrees) 
from single full-sky simulation.
We incorporate the masked regions associated with the positions of bright stars inside 
the HSC survey regions.
The details of the full-sky weak lensing simulations 
and masked regions are found in Paper I.
From 200 realizations of mock HSC data,
we construct smoothed convergence maps with the compensated Gaussian
filter and fully utilize these maps to derive the covariance matrices between
the statistics.

When calculating the inverse covariance, 
we include a debiasing correction as 
$\alpha=(n_{\rm real}-n_{\rm bin}-2)/(n_{\rm real}-1)$ 
with $n_{\rm rea}=200$ being the number of realization of simulation sets and
$n_{\rm bin}=26$ being the number of total bins in our data vector
\citep{Hartlap2007}.

We also take into account the constraints from the CMB priors 
expected from the Planck satellite mission.
When we compute the Fisher matrix for the CMB, we use 
the Markov-Chain Monte-Carlo (MCMC) engine for exploring cosmological 
parameter space {\tt COSMOMC} \citep{2002PhRvD..66j3511L}.
We consider the parameter constraints from the angular 
power spectra of temperature anisotropies, $E$-mode polarization and 
their cross-correlation.
For the MCMC analysis, in addition to $10^{9}A_{s}, n_{s}, \Omega_{c}h^2$ and $w_{0}$, 
our independent variables 
include the baryon density $\Omega_{b} h^2$, 
Hubble parameter $h$, 
reionization optical depth $\tau$, 
To examine the pure power of weak lensing statistics 
to constrain $w_{0}$ and $|f_{\rm R0}|$, 
we do not include any constraints on $w_{0}$ and $|f_{\rm R0}|$ from the CMB.
Assuming that the constraints from the CMB and the lensing statistics are independent, 
we express the total Fisher matrix as
\beqa
\bd{F} = \bd{F}_{\rm lensing} + \bd{F}_{\rm CMB}. \label{eq:Ftot}
\eeqa
When we include the CMB priors by Eq.~(\ref{eq:Ftot}), 
we marginalize over the other cosmological parameters 
except the following three parameters: $10^{9}A_{s}, n_{s}$, 
and $\Omega_{\rm m0}$.

\begin{figure*}
\centering \FigureFile(70mm,50mm){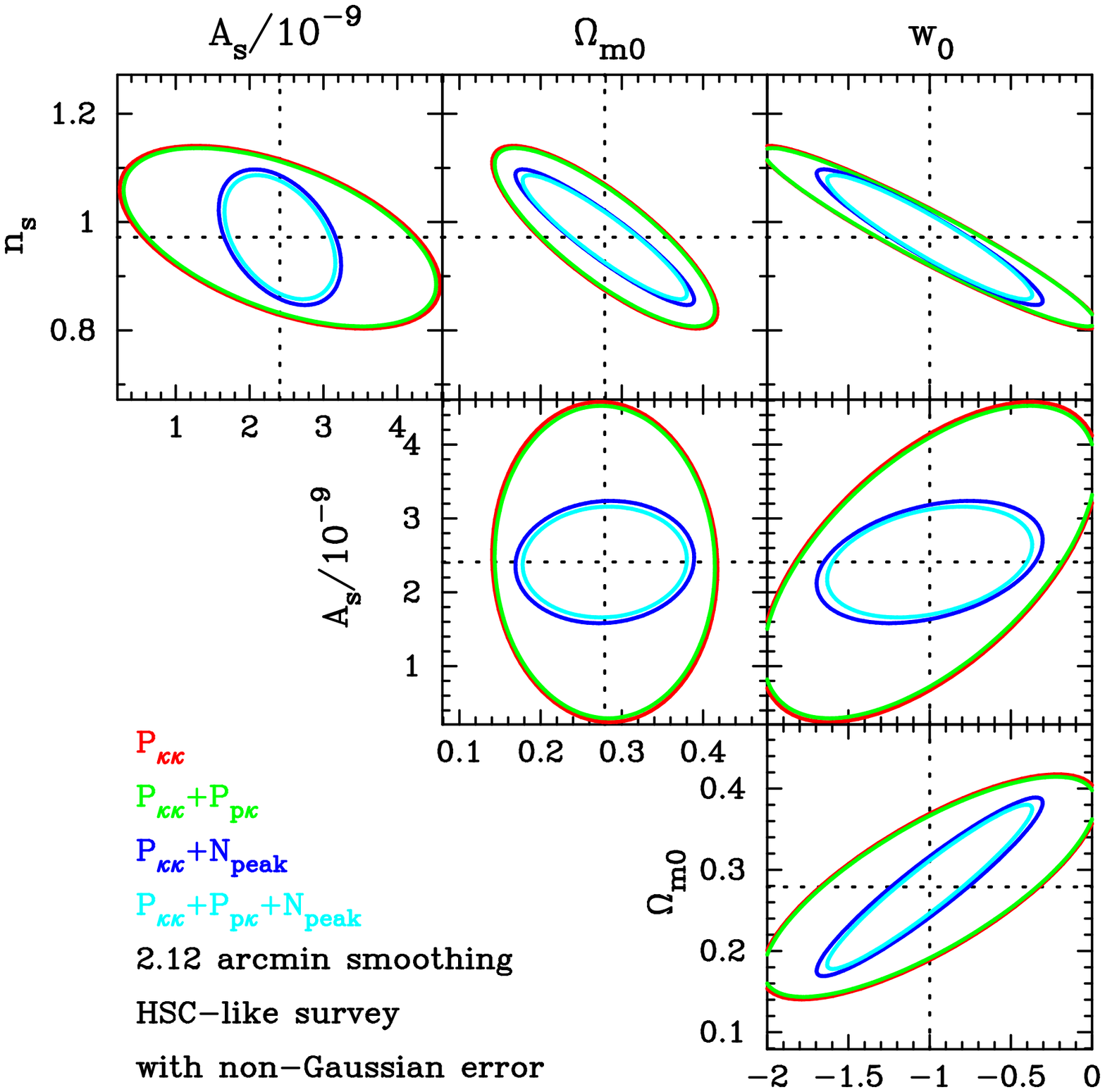}
\centering \FigureFile(70mm,50mm){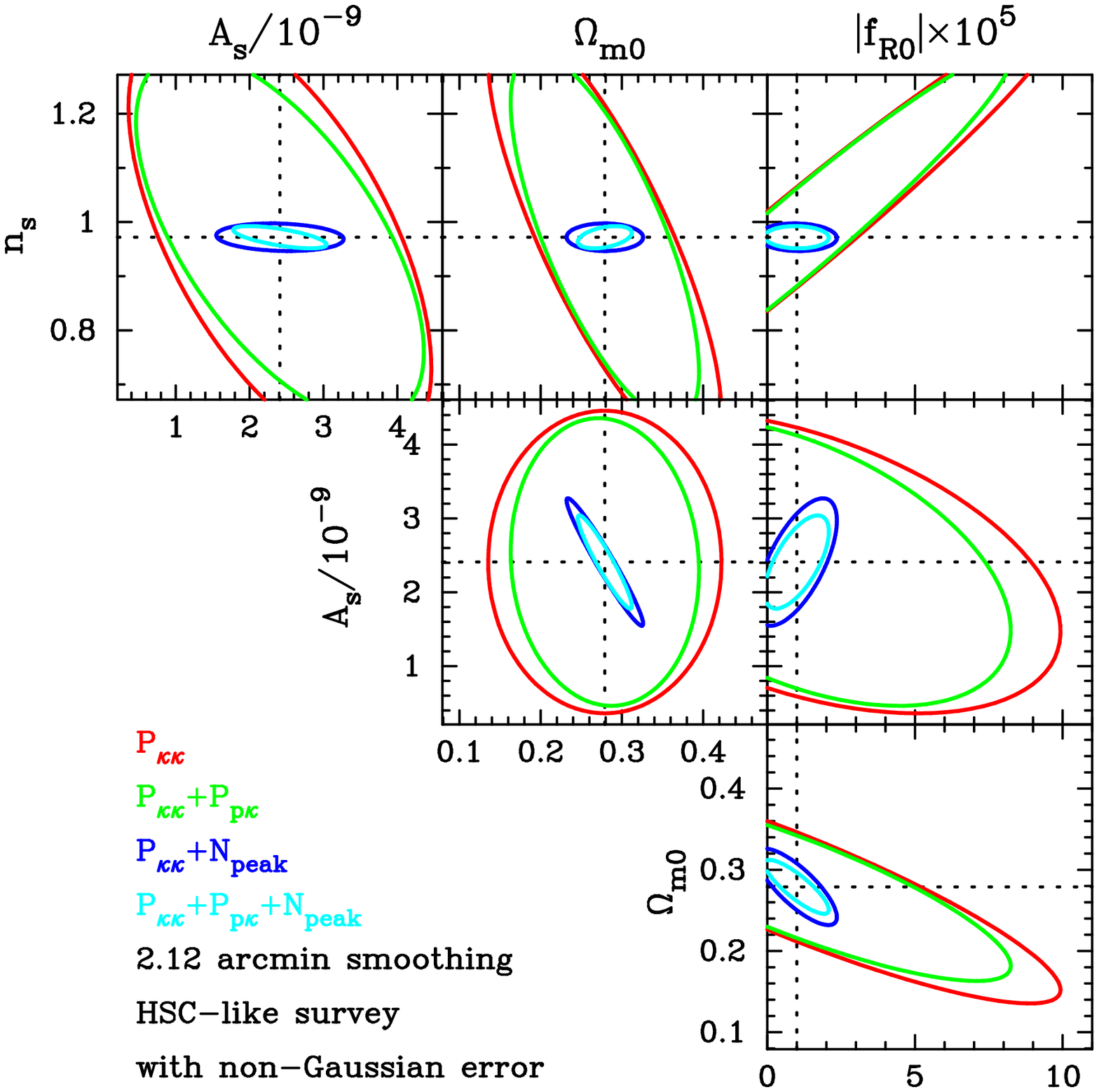}
\caption{
	The error circle obtained from Fisher analysis in the case of Hyper Suprime-Cam survey.
	In this figure, we show the cosmological constraints from weak lensing statistics alone.
	Each colored line corresponds to the constraints derived from different statistics:
	$P_{\kappa \kappa}$ (red),
	$P_{\kappa \kappa}$ and $P_{{\rm p}\kappa}$ (green),
	$P_{\kappa \kappa}$ and $N_{\rm peak}$ (blue), and
	combined with three (cyan).
	The left panel represents the constraint on $w$CDM model, while
	the right one shows the constraint on $f(R)$ model.	
	To estimate the constraints presented here,
	we take into account the non-Gaussian covariances derived from 200 masked sky simulations.  
	\label{fig:forecast_HSC}
	}
\end{figure*}

Figure \ref{fig:forecast_HSC} summarizes the expected constraints on two cosmological models
($w$CDM or $f(R)$) in the case of HSC survey.
The left panel shows the marginalized $1\sigma$ constraints on $w$CDM models
in two-dimensional parameter space
and the right panel shows the constraints on $f(R)$ models.
In each panel, we indicate constraints (error circles)
from different statistics by colored lines.

In the case of $w$CDM model, there remains strong degeneracy
among parameters;
especially, it is difficult to break the degeneracy
among $w_{0}$, $A_{s}$ and $\Omega_{\rm m0}$ 
with our statistics alone.
This is because we can measure the amplitude of matter density fluctuations by
using both $P_{\kappa \kappa}$ and $N_{\rm peak}$ and there apprear
no characteristic features in our lensing statistics.
In order to break such degeneracies, we require other information such as CMB measurements.
When including the information from CMB, we can successfully break the degeneracy among
cosmological parameters and thus measure $w_{0}$ with a level of $\sim$0.1.

The situation becomes slightly different in the case of $f(R)$ model.
The modification of gravitational force due to the additional scalar degree of freedom 
induces new characteristic features, which can not be compensated 
by other cosmological parameters.
These features are expected to show up as excess of the halo mass function 
at $M \sim10^{14}\, h^{-1}M_{\odot}$ \citep{2011PhRvD..84h4033L}, 
or excess in the lensing power spectrum at multipole $\ell\sim100-500$ \citep{2014ApJS..211...23Z}.
The relevant mass and length scales of the excess
are set by the value of $|f_{R0}|$.
Our weak lensing statistics presented here are sensitive and
thus one can measure such interesting features
using data from upcoming lensing surveys.
Interestingly, the expected constraint on $|f_{\rm R0}|$ is at a level of $10^{-5}$
with our weak lensing statistics \textit{alone}.  
In addition to our lensing statistics, parameter constraints
from CMB measurements would greatly help to
break degeneracies among other parameters than $|f_{\rm R0}|$ and $A_{\rm vir, 0}$.
Since the CMB information provides generally more stringent constraints
on $n_{s}$ and $A_{s}$,
we can further improve the constraint on $|f_{\rm R0}|$ 
by combining CMB 
with our lensing statistics.
The expected $1\sigma$ error for each parameter is summarized 
in Table \ref{tab:param_const_expect}.

\begin{table*}
\begin{center}
\begin{tabular}{|l|c|c|c|c|c|}
\hline
$w$CDM & $10^{9}A_{s}$ & $n_{s}$ & $\Omega_{\rm m0}$ & $A_{\rm vir, 0}$ & $w_{0}$ \\ \hline
$P_{\kappa\kappa}$ (HSC) 
& 1.43 (0.0409) & 0.111 (0.00385) & 0.0918 (0.00192) & -- (--) & 0.686 (0.0227) \\ \hline
$P_{\kappa\kappa}+P_{{\rm p}\kappa}$ (HSC) 
& 1.40 (0.0406) & 0.109 (0.00381) & 0.0897 (0.00190) & 1.77 (1.58) & 0.676 (0.0225) \\ \hline
$P_{\kappa\kappa}+N_{\rm peak}$ (HSC) 
& 0.548 (0.0338) & 0.0830 (0.00331) & 0.0728 (0.00170) & 1.14 (0.314) & 0.469 (0.0193) \\ \hline
$P_{\kappa\kappa}+P_{{\rm p}\kappa}+N_{\rm peak}$ (HSC) 
& 0.497 (0.0299) & 0.0758 (0.00291) & 0.0669 (0.00146) & 1.02 (0.306) & 0.417 (0.0165) \\ \hline
$P_{\kappa\kappa}+P_{{\rm p}\kappa}+N_{\rm peak}$ (HSC) + Planck 
& 0.132 (0.0294) & 0.0145 (0.00274) & 0.0124 (0.00143) & 0.379 (0.306) & 0.108 (0.0165) \\ \hline
\hline
$f(R)$ & $10^{9}A_{s}$ & $n_{s}$ & $\Omega_{\rm m0}$ & $A_{\rm vir, 0}$ & $10^{5}|f_{\rm R0}|$ \\ \hline
$P_{\kappa\kappa}$ (HSC) 
& 1.35 (0.0386) & 0.268 (0.00774) & 0.0951 (0.00170) & -- (--) & 5.91 (0.180) \\ \hline
$P_{\kappa\kappa}+P_{{\rm p}\kappa}$ (HSC) 
& 1.29 (0.0384) & 0.224 (0.00771) & 0.0768 (0.00169) & 1.82 (1.65) & 4.79 (0.178) \\ \hline
$P_{\kappa\kappa}+N_{\rm peak}$ (HSC) 
& 0.571 (0.0309) & 0.0168 (0.00372) & 0.0311 (0.00146) & 1.46 (0.296) & 0.897 (0.129) \\ \hline
$P_{\kappa\kappa}+P_{{\rm p}\kappa}+N_{\rm peak}$ (HSC) 
& 0.418 (0.0267) & 0.0136 (0.00297) & 0.0219 (0.00133) & 0.913 (0.291) & 0.717 (0.127) \\ \hline
$P_{\kappa\kappa}+P_{{\rm p}\kappa}+N_{\rm peak}$ (HSC) + Planck 
& 0.0898 (0.0263) & 0.00848 (0.00278) & 0.00735 (0.00130) & 0.677 (0.291) & 0.481 (0.127) \\ \hline
\end{tabular}
\caption{
	The summary of the expected marginalized error with $1\sigma$ confidence level.
	In order to make the forecast of constraints in this table,
	we assume the following parameters: 
	the rms of intrinsic ellipticies of sources $\sigma_{\gamma}=0.4$, 
	number density of sources $n_{\rm gal}=10\, {\rm arcmin}^{-2}$, 
	and source redshift $z_{\rm source}=1$.
	For comparison, we also provide the unmarginalized error 
	as the number in brackets.
	}
\label{tab:param_const_expect}
\end{center}
\end{table*}

\section{CONCLUSION AND DISCUSSION}\label{sec:con}

\begin{table*}
\begin{center}
\begin{tabular}{|l|c|c|c|c|c|}
\hline
$w$CDM & $10^{9}A_{s}$ & $n_{s}$ & $\Omega_{\rm m0}$ & $A_{\rm vir, 0}$ & $w_{0}$ \\ \hline
$P_{\kappa\kappa}$ (LSST) 
& 0.359 & 0.0279 & 0.0229 & -- & 0.171 \\
& (0.0102) & (9.64$\times10^{-4}$) & (4.80$\times10^{-4}$) & (--) & (0.00568) \\ \hline
$P_{\kappa\kappa}+P_{{\rm p}\kappa}$ (LSST) 
& 0.350 & 0.0272 & 0.0224 & 0.443 & 0.169 \\
& (0.0101) & (9.53$\times10^{-4}$) & (4.76$\times10^{-4}$) & (0.395) & (0.00564) \\ \hline
$P_{\kappa\kappa}+N_{\rm peak}$ (LSST) 
& 0.137 & 0.0207 & 0.0182 & 0.286 & 0.115 \\
& (0.00845) & (8.29$\times10^{-4}$) & (4.26$\times10^{-4}$) & (0.0785) & (0.00483) \\ \hline
$P_{\kappa\kappa}+P_{{\rm p}\kappa}+N_{\rm peak}$ (LSST) 
& 0.124 & 0.0189 & 0.0167 & 0.256 & 0.104 \\
& (0.00748) & (7.28$\times10^{-4}$) & (3.67$\times10^{-4}$) &  (0.0767) & (0.00414) \\ \hline
$P_{\kappa\kappa}+P_{{\rm p}\kappa}+N_{\rm peak}$ (LSST) + Planck 
& 0.0818 & 0.0115 & 0.0101 & 0.169 & 0.0665 \\
& (0.00747) & (7.25$\times10^{-4}$) & (3.66$\times10^{-4}$) &  (0.0767) & (0.00414) \\ \hline
\hline
$f(R)$ & $10^{9}A_{s}$ & $n_{s}$ & $\Omega_{\rm m0}$ & $A_{\rm vir, 0}$ & $10^{5}|f_{\rm R0}|$ \\ \hline
$P_{\kappa\kappa}$ (LSST) 
& 0.339 & 0.0671 & 0.0237 & -- & 1.47 \\
& (0.00965) & (19.3$\times10^{-4}$) & (4.26$\times10^{-4}$) & (--) & (0.0452) \\ \hline
$P_{\kappa\kappa}+P_{{\rm p}\kappa}$ (LSST) 
& 0.322 & 0.0562 & 0.0192 & 0.455 & 1.19 \\
& (0.00960) & (19.2$\times10^{-4}$) & (4.22$\times10^{-4}$) & (0.414) & (0.0447) \\ \hline
$P_{\kappa\kappa}+N_{\rm peak}$ (LSST) 
& 0.142 & 0.00422 & 0.00779 & 0.366 & 0.224 \\
& (0.00773) & (9.30$\times10^{-4}$) & (3.65$\times10^{-4}$) & (0.0741) & (0.0324) \\ \hline
$P_{\kappa\kappa}+P_{{\rm p}\kappa}+N_{\rm peak}$ (LSST) 
& 0.104 & 0.00340 & 0.00586 & 0.228 & 0.179 \\
& (0.00668) & (7.41$\times10^{-4}$) & (3.32$\times10^{-4}$) & (0.0729) & (0.0319) \\ \hline
$P_{\kappa\kappa}+P_{{\rm p}\kappa}+N_{\rm peak}$ (LSST) + Planck 
& 0.0655 & 0.00285 & 0.00369 & 0.204 & 0.151 \\
& (0.00667) & (7.38$\times10^{-4}$) & (3.32$\times10^{-4}$) & (0.0729) & (0.0319) \\ \hline
\end{tabular}
\caption{
	The expected marginalized $1\sigma$ error 
	in a LSST-like survey with sky coverage of 20,000 square degrees.
	We adopt the same parameter as shown in 
	Table~\ref{tab:param_const_expect}.
	We scale the covariance matrices 
	with survey area (i.e. by a factor of $20000/1250$).
	As for Table~\ref{tab:param_const_expect},
	the unmarginalized error is shown as the number in brackets.
	}
\label{tab:param_const_expect2}
\end{center}
\end{table*}

We have studied cosmological information content 
in statistics of weak-lensing selected clusters.
We have developed a theoretical model of the lensing statistics
and applied to two competing cosmological models: $w$CDM model and $f(R)$ model.
The key parameter in $w$CDM model and $f(R)$ model are 
the equation of state parameter of dark energy $w_{0}$
and the additional scalar degree of freedom $|f_{\rm R0}|$, respectively.
We have utilized masked sky simulations (see Paper I for details) 
to estimate non-Gaussian covariance caused by
non-linear gravitational growth and 
the mode-coupling due to masked regions simultaneously.
With such non-Gaussian covariance,
we have performed a Fisher analysis, which yields realistic forecast for constraining 
the nature of dark energy or the modification of gravity.

We consider specifically ongoing Hyper Suprime-Cam (HSC) survey 
with a sky coverage of $\sim 1250$ squared degrees.
Combined analysis of cosmic shear and weak lensing selected clusters 
can constrain $w_{0}$ with a level of $\Delta w_{0} \sim 0.1$, 
and $|f_{\rm R0}|$ with a level of $\Delta |f_{\rm R0}| \sim 5\times10^{-6}$
with the help of cosmic microwave background measurements.
Note that the expected constraint on $|f_{\rm R0}|$ is comparable to
the recent constraints of \citep{2014arXiv1412.0133C}.
Clearly, our approach is promising for upcoming surveys 
with a sky coverage of $10,000$ squared degrees or more. 
Assuming that the statistical error in upcoming wide field surveys is 
reduced proportionally to the effective survey area,
we can improve the constraints by a factor of $\sim (20000/1250)^{1/2} =4$
in the case of the Large Synoptic Survey Telescope (LSST)\footnote{
\rm{http://www.lsst.org/lsst/}
} 
with a proposed sky coverage of 20,000 square degrees.
We summarize the expected error with a LSST-like survey in Table~\ref{tab:param_const_expect2}.

\begin{figure}
\centering \FigureFile(70mm,50mm){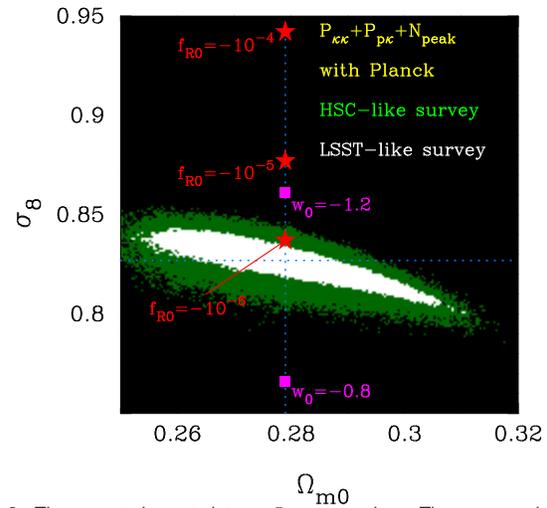}
\caption{
	The expected constraints on $\Omega_{\rm m0}-\sigma_{8}$ plane.
	The green and white region correspond to the marginalized 
	$95\%$ confidence level for HSC and LSST-like survey, respectively.
	We evaluate these regions from the Fisher matrix in combined analysis of 
	cosmic shear power spectrum, the cross correlation of convergence and its peak,
	and the number count of peaks with prior information of CMB measurement.
	The center of confidence surface represent the standard $\Lambda$CDM model, 
	while square and star symbols show the different value of $\sigma_{8}$
	in $w$CDM model and $f(R)$ model. 
	\label{fig:om_sigma8_forecast}
	}
\end{figure}

The normalization of power spectrum of matter density perturbations is 
one of the basic quantities in measure of gravitational growth at low redshift.
In practice, we normalize the linear matter power spectrum
by using the following quantities:
\beqa
\sigma_{R}(z) = \left(\int \frac{{\rm d}^3k}{(2\pi)^3} |W_{\rm TH}(kR)|^2 P^{L}_{m}(k, z)\right)^{1/2}, 
\label{eq:sigma8}
\eeqa
where $W_{TH}(kR)$ is the Fourier transform of top hat function with scale of $R$ 
and $R$ is commonly set to be 8 $\rm Mpc/h$.
Theoretically, this quantity $\sigma_{8}(z)$ can be ${\it derived}$ 
when the linear matter power amplitude is measured at higher redshift 
and the linear growth rate is taken into account properly.
On the other hand, 
we can {\it measure} the value of $\sigma_{8}(z)$ in a more direct manner
when we probe the clustering of matter density field at lower redshift.
Therefore, comparing the value of $\sigma_{8}$ 
derived from high redshift information with the observed $\sigma_{8}$ at lower redshift 
provides an invaluable opportunity to check consistency of cosmological models 
between high and low redshift \citep{2014arXiv1408.4742M}.
Figure \ref{fig:om_sigma8_forecast} shows
the expected constraint on $\sigma_{8}$
with the lensing statistics proposed in this paper.
In this figure, we present the expected constrained region on 
$\Omega_{\rm m0}-\sigma_{8}(z=0)$ plane with $95\%$ confidence level.
In order to derive such regions, we sample 100,000 points in five
cosmological parameters space 
for $w$CDM model (i.e. $A_s, n_s, \Omega_{\rm m0}, A_{\rm vir, 0}$ and $w_{0}$) 
assuming Gaussian distribution function
with covariance estimated from Fisher analysis 
(see, Section~\ref{sec:forecast} for details).
Then, we calculate the value of $\sigma_{8}(z=0)$ at each point.
With prior information of CMB measurement, our statistics can
constrain on $\sigma_{8}(z=0)$
with a level of 0.025 and 0.01 for HSC-like and LSST-like survey,
respectively.
Interestingly, in the case of LSST-like survey, we find that
$\sigma_{8}(z=0)$ can be constrained with precision
of $\sim$0.025 even in absence of CMB information.
In Figure \ref{fig:om_sigma8_forecast}, 
the square and star symbols represent the derived
values of $\sigma_{8}$ for $w$CDM and $f(R)$ models 
for a given $A_{s}$.
For $A_{s}$ derived from CMB,
dark energy and modified gravity model
predict the different values of $\sigma_{8}$
from that of the standard $\Lambda$CDM model.
Thus, the constraint on $\sigma_{8}$ by our lensing
statistics serves as an important test
on the self-consistency of the standard $\Lambda$CDM
model through $z\sim1000$ to $z\sim0$.

Combined statistical analysis with 
weak lensing selected clusters and cosmic shear 
provides precise measurement of gravitational growth at low redshift,
and thus can place robust constraints on variant cosmological models.
However, in order to apply our method to real data set,
we need to consider systematic effects in detail.
One of the most important systematics is source redshift uncertainty.
Throughout this paper, we have assumed
all the source galaxies are located at the same redshift.
In reality, source galaxies are distributed
over a wide redshift range.
In order to examine the effect of source redshift uncertainty, 
we compare the case of $z_{\rm source}=0.9$ and $z_{\rm source}=1$ under 
$\Lambda$CDM cosmology.
The 10\% difference in the mean source redshift roughly causes
under/over-estimate of $\Omega_{\rm m0}$ with a level of $\sim0.01$ 
for cosmic shear power spectrum and cross power spectrum of
convergence and its peaks,
while the number count of lensing peaks is less sensitive.
Therefore, the mean source redshift needs to be calibrated 
with a level of 0.1 for HSC survey.
Clearly, it is important
to study the effect of photometric redshift uncertainty
in a more quantitative manner, using, for instance,
realistic mock galaxy catalogs.

The baryonic processes such as gas cooling and stellar feedback
cause bias in parameter estimation with our lensing statistics.
Previous studies
\citep{Semboloni2011, 2013MNRAS.434..148S, 2013PhRvD..87d3509Z}
indicate that the baryonic effect on matter clustering could 
change two-point statistics of cosmic shear at $\ell\sim1000$
by a few percent.
Since a 3\% difference of $P_{\kappa\kappa}$ roughly corresponds to 
$\Delta \Omega_{\rm m0}\sim 0.015$, the baryonic effect needs to be
accounted for in the case of LSST-like survey.
For peak statistics, we already take into account a sort of
the baryonic effect by considering the uncertainty of halo concentration.
We argue that the baryonic processes would cause minor effect on peak statistics
for high peaks, as \citet{2015arXiv150102055O} have examined.
Nevertheless, the detailed modeling of baryonic effect on 
weak lensing selected clusters will allow to derive
more robust constraint on cosmological models.

Massive neutrinos are expected to have a significant effect on cosmic
structure formation 
(e.g., \cite{1980PhRvL..45.1980B}).
According to \citet{2015arXiv150606313H},
massive neutrinos with $M_{\nu} = 0.2$ eV  
suppress the amplitude of convergence power spectrum at $\ell\sim1000$
by a factor of $\sim0.9$ (see Figure 1 and 2 in \citet{2015arXiv150606313H}).
In addition, \citet{2013JCAP...12..012C} have shown
massive neutrinos with $M_{\nu} = 0.2$ eV change the halo mass function
with the halo mass of $\sim10^{14}\,h^{-1}M_{\odot}$ by a factor of $\sim 0.7$.
Therefore, ignoring massive neutrinos can potentially 
cause the mis-estimation of $\sigma_{8}$ and/or $\Omega_{\rm m0}$.
Further studies on  the
impact of massive neutrinos on weak lensing statistics
are warranted.

Magnification by weak lensing causes 
the scatter of the brightness and/or the size of source galaxies.
This magnification effect on cosmic shear statistics has been investigated
in literature \citep{2009ApJ...702..593S, 2011ApJ...735..119S, 2014PhRvD..89b3515L}.
For convergence power spectrum, 
\citet{2009ApJ...702..593S} have estimated that
the magnification effect typically induces $\sim$1\% differences of 
$P_{\kappa\kappa}$ at $\ell \sim 1000$.
On the other hand, the peak height by the NFW dark matter halo
is less sensitive to the magnification effect even if 
one reconstructs the lensing mass map with a smoothing scale of 
a few arcminutes (e.g., see Figure~5 in \citet{2011ApJ...735..119S}).
The magnification effect on peak statistics 
would be important in the case of LSST-like survey 
as \citet{2014PhRvD..89b3515L} pointed out.

Other than magnification effect,
there exist possible systematic effects 
which could be critical in cosmological analysis with weak lensing
selected clusters.
Source-lens clustering \citep{2002MNRAS.330..365H}
and the intrinsic alignment \citep{2004PhRvD..70f3526H}
are likely to compromise cosmological parameter estimation.
Although the impact of those effects on cosmic shear statistics is still uncertain,
a promising approach in theoretical studies 
would be associating the source positions with their host dark matter halos 
on the light cone. 

Future galaxy imaging surveys will generate a large amount of 
high-precision cosmic shear data. It will then become possible
to map the matter density distribution in the universe
and to measure the growth of the large-scale structure
in a direct manner.
The lensing statistics studied in this paper will enable us 
to make the best use of data from future large surveys
to explore the origin of cosmic acceleration, 
or the deviation from the concordance $\Lambda$CDM cosmology.



\section*{acknowledgments}
This work is supported in part by Grant-in-Aid for
Scientific Research from the JSPS Promotion of Science
(25287050; 26400285).
M.S. is supported by Research Fellowships of the Japan Society for 
the Promotion of Science (JSPS) for Young Scientists.
NY acknowledges financial support from JST CREST. 
Numerical computations presented in this paper were in part carried out
on the general-purpose PC farm at Center for Computational Astrophysics,
CfCA, of National Astronomical Observatory of Japan.



\end{document}